\documentclass[a4paper,12pt]{article}
\usepackage{graphicx}
 \usepackage{amsmath}
 \usepackage{amsfonts}
 \usepackage{mathtext}
 
\begin{document}

\begin{center}

{\Large \bf  Is  bimodality a sufficient condition for a first order phase transition existence?}

\vspace*{1.cm}

{\bf  K. A. Bugaev$^1$, A. I. Ivanytskyi$^1$, V. V. Sagun$^1$ and  D. R. Oliinychenko$^2$}\\

\vspace*{1.cm}

{\small \it $^1$Bogolyubov Institute for Theoretical Physics,
National Academy of Sciences of Ukraine,}\\
{\small \it Metrologichna str. 14$^b$, Kiev-03680, Ukraine}\\
{\small \it $^2$Physical Engineering Training-and-Research Center, Institute of  Physics, 
National Academy of Sciences of Ukraine,  Acad.Vernadskoho Blvd. 36, Kiev-03680,  Ukraine}\\

\vspace*{.55cm}


\end{center}


\abstract{
Here we present two explicit counterexamples to the widely spread beliefs 
about  an exclusive role of bimodality as the first order phase transition  signal. 
On the basis of an exactly solvable statistical model generalizing the  statistical multifragmentation model  
we demonstrate  that the bimodal distributions can naturally  appear both  in  infinite and in finite systems without a phase transition. In the first  counterexample 
a bimodal distribution appears  in an infinite system at the supercritical temperatures due to the negative values of the surface tension coefficient. In the second counterexample we explicitly demonstrate that a bimodal fragment distribution appears  in a finite volume analog of  a gaseous phase. 
In contrast to the  statistical multifragmentation model,  the developed statistical  model  corresponds to the compressible nuclear liquid  with the tricritical endpoint  located at one third of the normal nuclear density.
The suggested parameterization of the liquid phase equation of state is consistent with the L. van Hove axioms of statistical mechanics and it does not lead  to an appearance of the non-monotonic isotherms in the macroscopic mixed phase region which are typical for the classical models of the Van der Waals type. 
Peculiarly,  such a  way to account for  the nuclear liquid compressibility automatically leads to   an appearance of  an  additional state that in many respects resembles the physical antinuclear matter.
\\

}

%
\section{Introduction}\label{secintro}

During the last decade the studies  of the nuclear liquid-gas phase transition (PT)  stimulated both theoretical and  experimental  interest to the bimodal distributions \cite{Bmodal:Chomaz01,Bmodal:Chomaz03,Bmodal:Gulm04, Bmodal:Gulm07, Bmodal:Lopez06, Bmodal:Indra06,Bmodal:Bruno08, Bmodal:Indra09}.   Moreover, some  theoretical arguments \cite{Bmodal:Chomaz01,Bmodal:Chomaz03, Bmodal:Gulm04, Bmodal:Gulm07}, although  obtained  approximately, which relate  the bimodal distribution of  a certain order parameter and  the location of the Yang-Lee zeros \cite{YangLee:52} in a complex fugacity plane   became so  popular  that  nowadays the  bimodality is considered as a signal of the first order PT in finite systems, whereas the opposite opinions \cite{Moretto:05, Bmodal:Anti05,FS:Bugaev07}  are, in fact, ignored.  The scheme connecting the bimodality and  the Yang-Lee zeros \cite{Bmodal:Chomaz03, Bmodal:Gulm07} is so abstract and general that the authors  failed even   to discuss the physical origin of the bimodal distribution. However, in our opinion this is a crucial point, since in  the nuclear physics experiments at   intermediate energies one cannot get the purely statistical distributions of any observable because the process of collision is a dynamical one and, hence,  we  cannot account for or extract the dynamical fluctuations of the initial conditions,  the fluctuations of the number of participating nucleons, or possible instabilities occurring during the course of the system expansion and/or freeze out. 
Moreover,  it is not evident that the observed bimodal distributions are not generated by the imposed experimental cuts.

The authors of these theoretical  scheme \cite{Bmodal:Chomaz03,Bmodal:Gulm04, Bmodal:Gulm07}  implicitly assumed  that the measured distributions and the  corresponding  partition function of  the dynamically evolving system produced in the nuclear reaction generated by the recipe of  \cite{Bmodal:Chomaz03,Bmodal:Gulm04, Bmodal:Gulm07}  do, indeed,  correspond to the  equilibrium partition function of the original physical system.  This assumption, however, cannot be justified  without having a complete  dynamical model which correctly describes the whole evolution of the system. Moreover,  even, if one is able to completely account for the whole dynamical aspects of  the system evolution and, thus, is able to extract the purely statistical distributions, then there is no guaranty that the suggested theoretical  scheme 
\cite{Bmodal:Chomaz03,Bmodal:Gulm04, Bmodal:Gulm07} will work without any additional  conditions.  For example, it is absolutely  unclear what one should  do, if the extracted  statistical distributions  do not correspond to the statistical ensemble of the physical system under consideration? For the macroscopic systems we do not have such a  problem, since for  the vast majority of systems all the statistical ensembles are equivalent and, hence,  one can easily change them and choose the appropriate  one.  This, however, is not the case for finite or even small systems which are studied in the nuclear physics experiments.
 
The second typical mistake of Ref. \cite{Bmodal:Chomaz01,Bmodal:Chomaz03,Bmodal:Gulm04, Bmodal:Gulm07} and the similar schemes \cite{THill:1,DGross:1}  is  that the    authors of such schemes  identify each  local maximum of the bimodal distribution with a pure phase. Even in a famous textbook of T. Hill on thermodynamics of small systems  \cite{THill:1}  such an assumption is a corner stone of  his treatment  of PTs in finite systems.  In contrast to the authors of the scheme \cite{Bmodal:Chomaz03, Bmodal:Gulm04, Bmodal:Gulm07} Hill 
justified  his assumption on bimodality by stating that due to the fact that an interface between two pure phases 'costs' some additional energy, the probability of their coexisting 
in a finite system is less than for each of pure phases. 
We, however, should remind that the assumption on the pure phases existence in small system is taken from the examples of infinite systems, whereas for finite systems such an assumption cannot be justified. Moreover, the examples of the constrained statistical multifragmentation model (CSMM) \cite{Bugaev:CSMM05} and  the gas of hadron bags model \cite{FS:Bugaev07} which are exactly solved for finite systems  and  which allow one to rigorously define analogs of phases for finite grand canonical systems, show that, in contrast, to assumptions of Refs.  \cite{Bmodal:Chomaz01,Bmodal:Chomaz03, Bmodal:Gulm07, THill:1,DGross:1}, in finite systems the pure liquid phase cannot exist at finite pressures. Instead, it can  appear only as a part of mixed phase which is represented by even number of  thermodynamically metastable  states \cite{Bugaev:CSMM05,FS:Bugaev07}. 

Therefore, here we would like to give some counterexamples to the claims of  Refs.  
\cite{Bmodal:Chomaz01,Bmodal:Chomaz03, Bmodal:Gulm07, THill:1} by considering the exact analytical solutions of the CSMM in the thermodynamic limit and for  the finite volumes which lead to the bimodal fragment distributions inside of  the cross-over  region  and inside of the  gaseous phase. For this purpose we consider the CSMM with two new elements. The first of them is a more realistic equation of state for the liquid phase which, in contrast to the original SMM formulation \cite{SMM:Bondorf95, SMM:Bugaev00,Reuter:01}, is a compressible one. The second important element of the present model  is a more realistic  parameterization for the temperature dependence of surface tension that is  based on the exact analytical solution of the  partition function of surface deformations \cite{HDM:Bugaev05, HDM:Bugaev07}. Besides these two new elements allow us to study a realistic phase diagram of the CSMM  both for  finite systems and for infinite system. 

The work is organized as follows. In sect. 2 we describe the new parameterization of the CSMM liquid phase pressure which repairs the two main pitfalls of the original SMM and allows one to consider the compressible liquid which has the tricritical endpoint at the phase diagram at the one third of the normal nuclear density. It is also shown that the bimodal fragment size  distributions may appear at the supercritical temperatures  due to negative values of the surface tension coefficient and without any PT.  Sect. 3 is devoted to the analysis of  finite systems using  the exact solution of  CSMM. 
In this section we demonstrate that the bimodal fragment size  distribution is generated within the finite volume analog of the gaseous phase.  Our conclusions are formulated in sect. 4.


%
%
\section{CSMM with compressible nuclear liquid in thermodynamic limit}\label{secmodel}

The general solution of the CSMM  partition function formulated in the grand canonical variables of volume $V$, temperature $T$ and baryonic chemical potential $\mu$ is given by \cite{FS:Bugaev07, Bugaev:CSMM05,Bugaev:Thesis10,Bugaev:Nucleation11}
\begin{equation}\label{EqI}
{\cal Z}(V,T,\mu)~ = \sum_{\{\lambda _n\}}
e^{\textstyle  \lambda _n\, V }
{\textstyle \left[1 - \frac{\partial {\cal F}(V,\lambda _n)}{\partial \lambda _n} \right]^{-1} } \,,
\end{equation}
where  the set of  $\lambda_n$ $(n=0,1,2, 3,..)$ are all the complex roots of  the equation 
\begin{equation}\label{EqII}
\lambda _n~ = ~{\cal F}(V,\lambda _n)\,,
\end{equation}
ordered as   $Re(\lambda_n) > Re(\lambda_{n+1})$ and $Im (\lambda_0) = 0$. The function ${\cal F}(V,\lambda)$ is defined as 
\begin{eqnarray}\label{EqIII}
&&\hspace*{-0.04cm}{\cal F}(V,\lambda) = 
  \left(\frac{m T }{2 \pi} \right)^{\frac{3}{2} }  z_1
~ \exp \left\{ \frac{\mu- \lambda T b}{T}  \right\} + \hspace*{-0.1cm} \sum_{k=2}^{K(V) }
\phi_k (T) \exp \left\{  \frac{( p_l(T,\mu)- \lambda T)b k }{T} \right\} 
\,.
\end{eqnarray}
Here $m \simeq 940$ MeV is a nucleon mass, $z_1 = 4$ is an internal partition (the degeneracy factor) of nucleons, $b = 1/ \rho_0 $ is the eigen volume of one nucleon in a vacuum ($\rho_0\simeq 0.17$ fm$^3$ is the normal nuclear density at $T=0$ and zero pressure). The reduced distribution function of the $k$-nucleon fragment in (\ref{EqIII}) is defined as 
\begin{equation}\label{EqIV}
 \phi_{k>1} (T)  \equiv \left(\frac{m T }{2 \pi} \right)^{\frac{3}{2} }  k^{-\tau}\, \exp \left[ - \frac{\sigma (T)~ k^{\varsigma}}{T}  \right]\,,
\end{equation} 
where $\tau \simeq 1.825$ \cite{Reuter:01} is the Fisher topological exponent and $\sigma (T)$ is the $T$-dependent surface tension coefficient.  Usually, the constant, parameterizing the dimension of surface in terms of the volume is  $\varsigma = \frac{2}{3}$, but in this work we would like to give the results for a wide range of  its values, namely for  $0 < \varsigma < 1$.

In (\ref{EqIII}) the exponents $\exp( - \lambda b k)$ ($k=1,2,3,...$) appear due to the hard-core  repulsion between the nuclear fragments \cite{Bugaev:CSMM05,SMM:Bugaev00,Bugaev:Thesis10}, while $p_l(T,\mu)$ is the pressure of the liquid phase \cite{Bugaev:Thesis10, Bugaev:Nucleation11}. 
As one can see from (\ref{EqIII}) the nucleons are treated  differently compared to larger fragments:  they do not have  the surface free energy and  all the bulk free energy characteristics  except for the baryonic charge which are encoded in the liquid phase pressure $p_l(T,\mu)$ (see later).  In principle, the fragments with the  mass below ten nucleon masses can be parameterized in a similar way \cite{SMM:Bondorf95, Bugaev:Thesis10}, but for the sake of  simplicity we treat in this way the nucleons only. 
Such a treatment  does not affect the properties of the phase diagram in the thermodynamic limit, since
exclusion  of any finite number of light fragments from the sums in   (\ref{EqIII}) does not affect
the PT existence and its order \cite{Bugaev:CSMM05,SMM:Bugaev00,Bugaev:Thesis10}.

Note also  that the complex free energy density $-T {\cal F}(V,\lambda)$ \cite{Bugaev:CSMM05} of the present  model   contains neither  the Coulomb energy nor  the asymmetry energy. This assumption  is similar to  Refs. \cite{SMM:Bugaev00, Reuter:01, SMM:simple98} and allows  us to study the nuclear matter properties in the thermodynamic limit. However,  in contrast to Refs. \cite{SMM:Bugaev00, Reuter:01, SMM:simple98}, the model free energy density $-T {\cal F}(V,\lambda)$ in (\ref{EqIII}) contains the liquid phase pressure that can be chosen in a general form and 
the size of maximal fragment $K(V)$ that  can be a desired function of the system volume $V$. However, in this section we consider the thermodynamic limit only, i.e. for $V \rightarrow \infty$  it follows $K(V) \rightarrow \infty$. Then the treatment of the model is essentially simplified, since Eq. (\ref{EqII}) can have only two kinds of solutions \cite{SMM:Bugaev00,Bugaev:CSMM05,Bugaev:Thesis10}, either the gaseous pole $p_g (T, \mu) = T \lambda_0 (T, \mu)$ for  ${\cal F}(V,\lambda_0 - 0) < \infty$ or the liquid essential singularity 
$p_l (T, \mu) = T \lambda_0 (T, \mu)$ for  ${\cal F}(V,\lambda_0 - 0) \rightarrow  \infty$. The mathematical reason why  only the rightmost solution   $\lambda_0 (T, \mu) = \max \{Re(\lambda_n)\} $  of   Eq. (\ref{EqII}) 
 defines the system pressure is evident from Eq. (\ref{EqI}): in the limit $V \rightarrow \infty$ all  the solutions  of (\ref{EqII}) other than the rightmost one are exponentially suppressed.  

In the thermodynamic limit the model has a PT, when there occurs a change of the  rightmost solution type , i.e. when the gaseous pole is changed by  the liquid essential singularity or vice versa. The PT line $\mu = \mu_c (T)$ is a solution of  the equation of  `colliding singularities' $p_g (T, \mu) = p_l (T, \mu) $ 
\cite{SMM:Bugaev00,Bugaev:CSMM05,Bugaev:Thesis10}, which is just the Gibbs criterion of  phase equilibrium. The properties of a PT are defined only by the liquid phase pressure  $p_l (T, \mu)$ and   by the temperature dependence of  surface tension $\sigma(T)$, since the value of Fisher exponent $\tau = 1.825$ is fixed by  the values of the critical indices of ordinary liquids \cite{Reuter:01} and by the experimental findings \cite{ISIS:99, EOS:00}.
 
In order to avoid the incompressibility of the nuclear liquid we suggest to consider the following simplest parameterization of its pressure 
\begin{eqnarray}
\label{EqV}
p_l=\frac{ W(T) +  \mu + a_2 ( \mu -\mu_0)^{2} + a_4 ( \mu -\mu_0)^{4}}{b} \,.
\end{eqnarray}
Note that  the above way to account for  the nuclear liquid compressibility  is fully  consistent with the L. van Hove axioms of statistical mechanics  \cite{VanHove,VanHove2}  and, hence,  it does not lead to an appearance of the non-monotonic isotherms in the mixed phase region which are typical for the mean-field models.
In \cite{Bugaev:Nucleation11} the liquid phase pressure was parameterized as a second order polynomial in  
the baryonic chemical potential. In our mind  Eq.   (\ref{EqV}) is more favorable, since it  allows one to easily get a correct value for the nuclear incompressibility factor for a normal nuclear liquid. 
In Eq.  (\ref{EqV}) $ W(T) = W_0 + \frac{T^2}{W_0}$ denotes  the usual  temperature dependent  binding energy per nucleon   \cite{SMM:Bondorf95, SMM:Bugaev00} with $W_0 =  16$ MeV and  the constants  $\mu_0$, $a_2$ and   $a_4 >0$.  In principle, these constants  should  be  fixed 
in the way  to
reproduce  the properties  of normal nuclear matter, i.e. at vanishing temperature  $T=0$ and normal nuclear density $\rho = \rho_0$ the liquid pressure must be zero
\begin{eqnarray}
\label{EqVI}
W_0 +  \mu_c(0) +  a_2 ( \mu_c(0) -\mu_0)^{2} + a_4 ( \mu_c(0) -\mu_0)^{4}  = 0\, ,  
\end{eqnarray}
where $\mu_c(0)$ is the baryonic chemical potential at the PT line taken at $T=0$. Finding the particle density of the liquid as  $\rho_l=\frac{\partial p_l}{\partial\mu}$ 
\begin{eqnarray}
\label{EqVII}
\rho_{l} (\mu)=\frac{1+ 2\, a_2\tilde \mu + 4\, a_4 \, \tilde \mu^{3}}{b}\, ,  \quad {\rm with} \quad \tilde \mu = \mu - \mu_0  \, , 
\end{eqnarray}
one can get the equation  for $\mu_c(0)$, i.e. from $ \rho_l (\mu_c(0)) = \rho_0$ it follows 
$2\, a_2\tilde \mu(0) + 4\, a_4 \, \tilde \mu(0)^{3} =0$, where the shifted  chemical potential $\tilde \mu (0)$ is defined as $\tilde \mu (0) \equiv \mu_c (0) - \mu_0 $.  Usually, an additional requirement to fix the nuclear liquid model parameters is related to the  incompressibility factor  
 of  the normal nuclear matter  \cite{MyEOS:93}  which is defined as 
\begin{eqnarray}
\label{EqVIII}
K_0 \equiv 9  \left(  \frac{\partial p_l}{\partial \rho_l} \right)_{T=0} = \frac{9 (1+ 2\, a_2\tilde \mu(0) + 4\, a_4 \, \tilde \mu(0)^{3} )}{2\, a_2 + 12  \, a_{4}\,  \tilde\mu(0)^{2}}  \,. 
\end{eqnarray}
The present day experimental estimates for the  incompressibility factor are $K_0^{exp} = 230 \pm 30$ MeV 
\cite{Kfactor:1, Kfactor:2,Kfactor:3,Khan:2009}, but the models with the typical 
value $K_0 = 300-360 $ MeV are also well known \cite{MyEOS:93, Kfactor:3}. For instance,  the Skyrme force 
model SIII, which is able to successfully  describe the experimental properties of many 
nuclei \cite{Kfactor:3}, has the value of the nuclear incompressibility factor $K_0 = 355$ MeV.
Therefore, instead of describing exactly the present day values of the normal nuclear incompressibility factor and have many additional parameters, we prefer to keep the model as simple as possible, but to  require that at 
 the tricritical point  the baryonic density  is   $\rho_{cep} = \rho_0/3$ which  is typical  for the liquid-gas PTs \cite{Stanley:71}. The latter generates the following equation for the shifted value of the baryonic chemical potential at 
 the tricritical endpoint: $2\, a_2\tilde \mu_{cep} + 4\, a_4 \, \tilde \mu_{cep}^{3} = - \frac{2}{3}$, where 
 $\tilde \mu_{cep} \equiv  \mu_{cep} - \mu_0$. 
 
 Choosing $\mu_0 = - W_0 = - 16$ MeV, we obtain $\tilde \mu (0) = 0$ and, hence, the expressions  (\ref{EqVI})
 and (\ref{EqVII}) are essentially simplified, respectively,  giving us $ \rho_l (\mu_c(0)) \equiv  \rho_0$  and 
 $K_0 = \frac{9}{2 \, a_2}$. Then,   solving  the phase equilibrium condition at the tricritical endpoint together with the condition on the baryonic density at  this point,  one can express both the coefficient $a_4$ and $\tilde \mu_{cep}$ in terms of $a_2$ and the pressure of gaseous phase $p_g (T_{cep}, \mu_{cep})$ taken at this point. Thus, one can express $K_0$,  $\rho_l (\mu_c(T_{cep})) $ and $a_4$ in terms of $a_2$ and $p_g (T_{cep}, \mu_{cep})$. However, we found that for 
$K_0 < 350$ MeV the obtained values of the coefficient $a_4$ are negative which leads to an  instability of 
 nuclear liquid at very high baryonic densities.  Therefore, in order to avoid these problems, we 
fixed $K_0 = 365$ MeV which leads to $a_2 \simeq 1.233 \cdot 10^{-2}$ MeV$^{-1}$ and $a_4 \simeq 4.099 \cdot 10^{-7}$ MeV$^{-3}$. Thus,   the present model  is able to repair the two major  unrealistic features of the original SMM, namely, it provides one with a reasonable value  for  the nuclear    liquid compressibility and with 
a physically motivated value for the baryonic density at the tricritical endpoint.

In addition to the new parameterization of the free energy of the $k$-nucleon fragment (\ref{EqIII}) we  propose to consider a more general parameterization of the surface tension coefficient 
\begin{equation}\label{EqIX}
 \sigma (T) =  \sigma_0 \left| \frac{T_{cep} - T }{T_{cep}} \right|^\zeta  {\rm sign} ( T_{cep} - T) ~,
\end{equation}
with  $\zeta = const \ge 1$, $T_{cep} =18$ MeV and $\sigma_0 = 18$ MeV the SMM 
\cite{SMM:Bondorf95}. In contrast to the Fisher droplet model \cite{Fisher:67} and   the usual SMM \cite{SMM:Bondorf95}, the CSMM surface tension (\ref{EqIX}) is negative above the critical temperature $T_{cep}$. It is necessary to stress that there is nothing wrong 
or unphysical with the negative values of surface tension coefficient  (\ref{EqIX}), since $ \sigma (T)\, k^\varsigma$ in (\ref{EqIV}) is the surface  free energy of the
fragment  of mean volume $b \,k $ and, hence, as any free energy,  it contains the energy part
$e_{surf}$ and  the entropy part $s_{surf}$ multiplied by temperature $T$ \cite{Fisher:67}. Therefore, at low temperatures the energy part dominates and the  surface  free energy is positive, whereas at high temperatures the number of fragment  configurations with large surface  drastically increases  and it exceeds  the  Boltzmann suppression and, hence,
the surface free energy becomes negative since $s_{surf} > \frac{e_{surf}}{T}$.
Because of this reason   the negative values of the surface tension coefficient  were recently  employed in a variety of exactly solvable statistical models for the deconfinement PT 
\cite{QGBSTM1,QGBSTM2,FWM:08,Aleksei:11}. For the first time  this fact was  derived  within the  exactly solvable model for surface deformations of large physical clusters  \cite{HDM:Bugaev05}. Very recently 
 an important  relation between the surface tension of large quark gluon bags and the string tension  of two  static  color charges measured by the lattice QCD  was  derived \cite{String:10}. Based on such a  relation  it was possible to conclude that at high temperatures the surface tension coefficient of quark gluon bags should be negative \cite{String:10,String:11}.

\begin{figure}[t]
%
%
\centerline{
\includegraphics[height=11.11 cm]{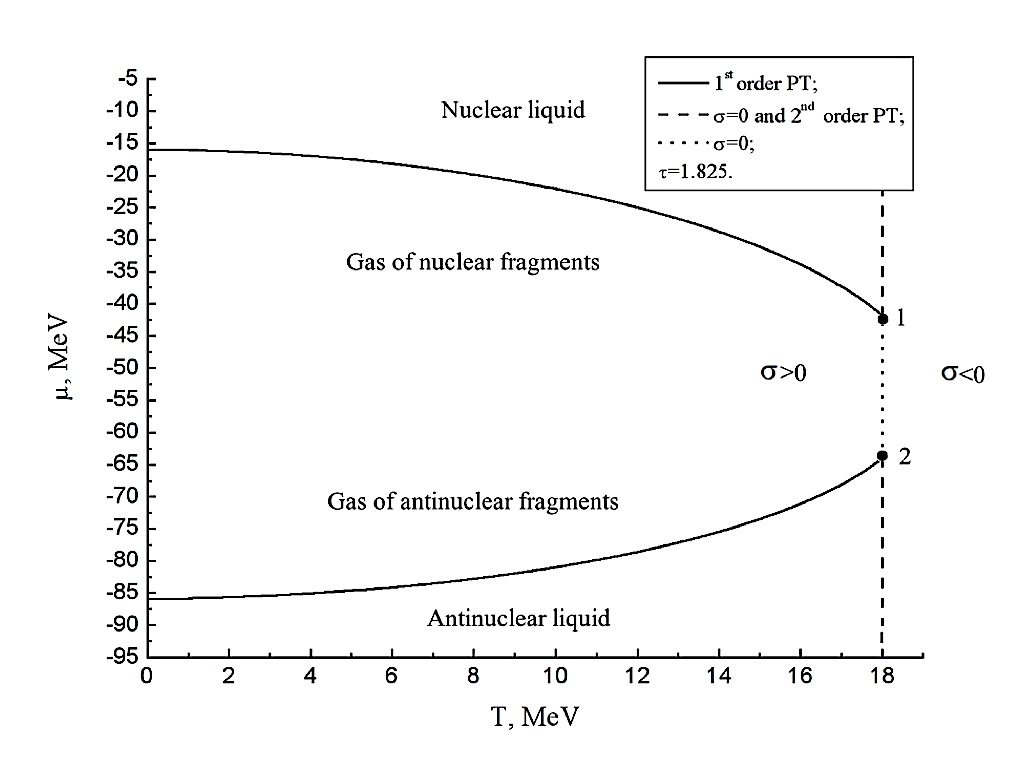} 
}  
 \caption{The phase diagram in $T-\mu$ plane. The first order PT occurs along the solid curves. Above the upper curve there exists the nuclear matter, while below the lower one there is  an analog of the antinuclear matter. The vertical dashed lines show the second order PT and the black circles correspond to the tricritical endpoints marked by the digits 1 (nuclear matter) and 2 (antinuclear matter). A cross-over occurs along the  dotted vertical line of the vanishing surface tension coefficient.   
}
  \label{fig1}
\end{figure}

Furthermore,  a thorough  analysis of the temperature dependence of the surface tension coefficient  in ordinary liquids \cite{ KABScaling:06,KABJGross:09} shows not only  that 
the surface tension coefficient approaches zero, but, in contrast to the widely  spread beliefs, for many liquids  the  full $T$ derivative of $\sigma  (T)$  does not vanish and remains finite at  $T_{cep}$: $\frac{d~ \sigma  (T)}{d ~T} < 0$ \cite{KABScaling:06}. Therefore, just the naive extension of these data to the temperatures above  $T_{cep}$ would lead to  negative values of  surface tension coefficient at the supercritical temperatures. On the other hand, if one, as usually,   believes that $\sigma \equiv 0$ for $T >T_{cep} $, then  it is absolutely unclear what  physical process can lead to  simultaneous existence of  the discontinuity of $\frac{d~ \sigma}{d ~T}$ at 
$T_{cep}$ and the  smooth behavior   of the pressure's  first and second derivatives   at the cross-over. 
Finally, the negative values of the surface tension at supercritical temperatures is the  only 
known physical reason which prevents the condensation of smaller droplets into a liquid phase and, thus, it terminates the first order  PT existence and degenerates it into a cross-over at these temperatures. 
Therefore, we conclude that  negative values of the surface tension coefficient   at supercritical temperatures are  also necessary for ordinary liquids although up to now this question has not been  investigated.

\begin{figure}[t]
\centerline{
\includegraphics[height=11.11 cm]{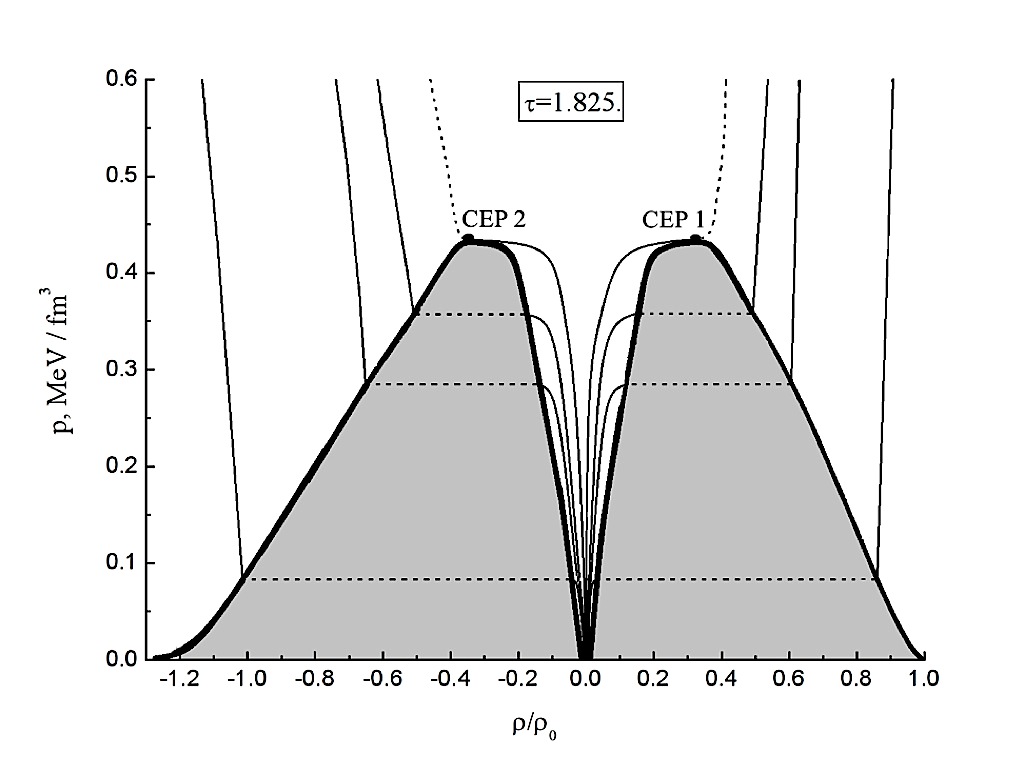} 
}  
%
 \caption{
The phase diagram in $\rho-p$ plane. The grey  areas show the mixed phases of the first  order PTs. 
The isotherms are shown for $T=11, 16, 17, 18$ MeV  from bottom to top. 
Negative density values correspond   to an   `antimatter'.
  For the densities  $|\rho/\rho_0| \ge 1/3 $ at    the isotherm $T = 18$ MeV there exists the second order PT. The tricritical endpoints are marked by the digits 1 (nuclear matter) and 2 (antinuclear matter). 
}
  \label{fig2}
\end{figure}

Similarly to the  simplified SMM  \cite{SMM:Bugaev00, Reuter:01},  for  $T < T_{cep}$  the present model has the nuclear liquid-gas PT of the first order. However, as one can see from Fig. \ref{fig1} in this region of temperatures  the model has two  first order PTs. The meaning of the second PT curve can be understood from 
Fig. \ref{fig2}. At first glance a mathematical cause of  an `antimatter'  appearance may look surprising since the gas pressure contains no fragments with negative baryonic charges. However,  this is true for  $\left| \frac{\tilde \mu}{T} \right| \ll 1$ only, while 
for   $\left| \frac{\tilde \mu}{T} \right| \ge   1$ the main contribution in the liquid phase pressure $p_l$  in 
(\ref{EqV}) is defined 
by the term $a_4 \tilde \mu^4$ and, hence its derivative with respect to $\mu$ determines a sign of  the baryonic charge 
density of  both a liquid phase and  a gas of  nuclear fragments. The letter can be  seen from the charge density expression for  the gaseous phase.  Indeed, finding  the $\mu$ derivative  of the gaseous phase pressure  $p_g = T \lambda_0 (T, \mu)$ from Eqs. (\ref{EqII}) and Eqs. (\ref{EqIII}), one finds the baryonic charge  density of  the gaseous phase as 
\begin{eqnarray}\label{EqX}
&&\hspace*{-0.04cm}\rho_g  = \frac{
\rho_0\, \left(\frac{m T }{2 \pi} \right)^{\frac{3}{2} }  z_1
~ \exp \left\{ \frac{\mu- \lambda T b}{T}  \right\} +  \rho_l \, \sum_{k=2}^{\infty }
\phi_k (T) \, k\, \exp \left\{  \frac{( p_l(T,\mu)- p_g(T,\mu))b k }{T} \right\} }{1 +
 \left(\frac{m T }{2 \pi} \right)^{\frac{3}{2} }  z_1
~ \exp \left\{ \frac{\mu- \lambda T b}{T}  \right\} +  \sum_{k=2}^{\infty }
\phi_k (T) \, k\, \exp \left\{  \frac{( p_l(T,\mu)- p_g(T,\mu) )b k }{T} \right\} 
} \,.
\end{eqnarray}
From this expression one can see that, if the contribution of the nucleons (proportional to $z_1$) is small compared to the sum over other nuclear fragments, i.e. for $\left(\mu - b\, p_l(T,\mu) \right)/T <  -1$, then 
the  baryonic charge  density of  the gaseous phase is proportional to that one of liquid, i.e. 
${\rm sign} \left[ \rho_g  \right]=  {\rm sign} \left[ \rho_l \right]$. 
Of course, one should not take this additional solution as a physical  antinuclear matter, since the gas pressure  of the present model  contains  only the nuclear fragments with the charges $k =1, 2, 3, ...$ that stay in front of  the nonrelativistic value  of the baryonic chemical potential $\mu$ and does not contain any terms with an opposite value of  $\mu$. It is clear  that  in a 
relativistic treatment one would have the symmetry with  respect to the charge conjugation $\mu_{rel} \leftrightarrow  -\mu_{rel}$ for the relativistic baryonic chemical potential $\mu_{rel} \equiv m + \mu$. Nevertheless, it is a remarkable fact, that  
the simplest way to account for  the nuclear liquid compressibility  which  is consistent with the L. van Hove axioms of statistical mechanics  \cite{VanHove,VanHove2} automatically leads to   an appearance of  an  additional state that in many respects resembles the physical antinuclear matter. 

\begin{figure}[t]
\centerline{
\includegraphics[height=11.11 cm]{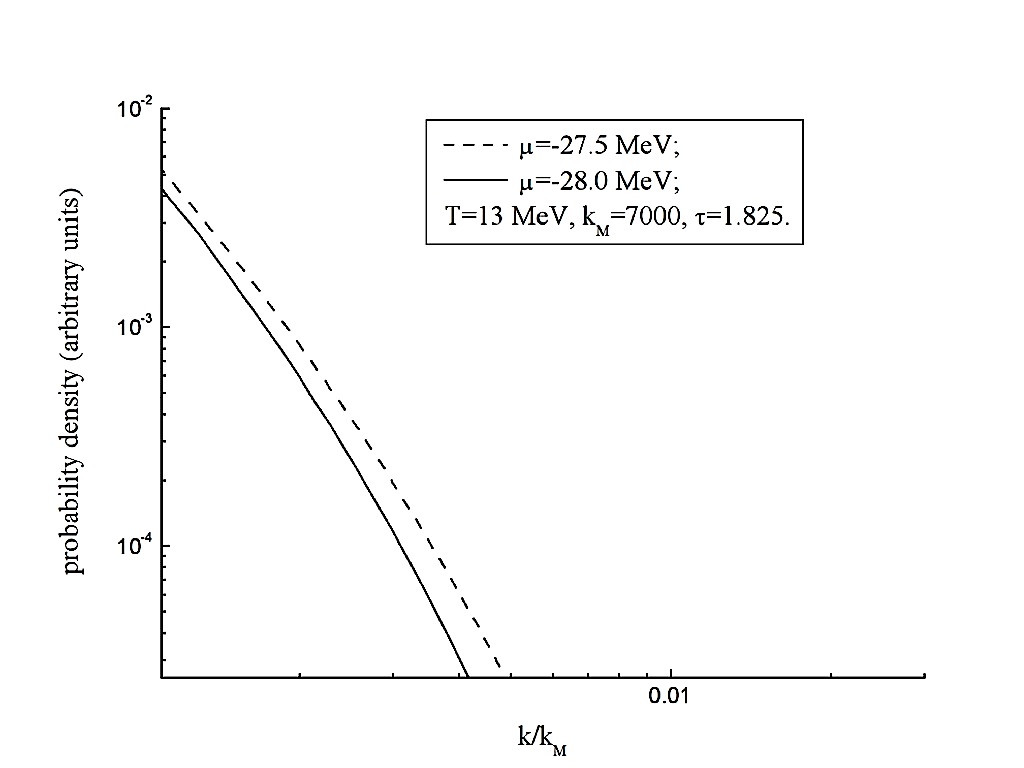} 
}  
 \caption{Fragment size distribution in the gaseous phase  is shown for a fixed temperature $T = 13$ MeV and two values of the  baryonic chemical potential $\mu$. The number of nucleons in  a fragment is  $k$.  The larger value of  $\mu$ corresponds to the gaseous state at the phase boundary with the mixed phase. The calculations were made for the largest fragment of $K(V) = k_{M}= 7000$ nucleons. 
}
  \label{fig3}
\end{figure}

Also Eq. (\ref{EqX}) clearly shows that at the phase equilibrium, i.e. for the same pressure, the  baryonic densities of  gaseous and liquid phases differ, if the sum staying both in numerator and in denominator of  (\ref{EqX})  is not divergent. This is possible,  either for positive values of the surface tension coefficient $\sigma(T) > 0$ and any positive value $\tau >0 $ or,
alternatively, for  $\sigma(T) = 0$ and  $\tau > 2$.  In either of these two cases there is a first order PT. If, however,  $\sigma(T) = 0$ and  $\tau \le  2$, which is the case for the present model at  $T=T_{cep}$, then for some values of the chemical potential one has  $\rho_g (T_{cep}, \mu_{cep}) = \rho_l (T_{cep}, \mu_{cep}) $ and  the sums in  (\ref{EqX}) diverge. Then at these points  there exists a PT of higher order.   The analysis of higher order derivatives of gaseous pressure made similarly to \cite{QGBSTM1}  shows that for $2 \ge \tau > \frac{3}{2}$ at the critical endpoint  of this model  there exists a second order PT. 
In the present model a second order PT exists not only at  the critical endpoints, but at the  two lines in the $T-\mu$ plane along which  the surface tension is zero (see the two vertical dashed lines  in Fig. \ref{fig1}).  Therefore, the both critical endpoints of the present model are the tricritical endpoints. 
This feature is similar to the simplified SMM \cite{SMM:Bugaev00, Reuter:01} and it is robust for $\tau =1.825$, whereas as one can see from Fig. \ref{fig2} the second order PTs of this model  are  not located  at the constant density as in the simplified SMM.
Finally, for the supercritical temperatures the surface tension (\ref{EqIX}) is negative and, hence, the phase equilibrium is not possible in this case \cite{SMM:Bugaev00, Reuter:01,Bugaev:Thesis10}.

Now we would like to study the fragment size distribution in two regions  of the phase diagram in order to elucidate the role of the negative surface tension coefficient.  In order to demonstrate the pitfalls of the bimodal concept of Refs. \cite{Bmodal:Chomaz01,Bmodal:Chomaz03, Bmodal:Gulm04, Bmodal:Gulm07,THill:1} we study only the gaseous phase and the supercritical temperature region, where there is  no  PT  by construction.  As one can see from Fig. (\ref{fig3}) in the gaseous phase, even at the boundary with the mixed phase,  the size distribution is a monotonically   decreasing function of  the number of nucleons in a fragment $k$. 
The found  distributions  are very similar to those one shown  in Fig. 5 of  \cite{SMM:12} for comparable temperatures.
As one can see from Fig. \ref{fig3} for  small fragments the  distribution is almost 
 power-like one (notice  the double logarithmic scale in Fig. \ref{fig3}), while for larger fragments the  deviation from a pure power law is seen. No bimodal distribution is found in this case, although in actual simulations we used $K(V) = k_{M}= 7000$ nucleons.

\begin{figure}[t]
\centerline{
\includegraphics[height=11.11 cm]{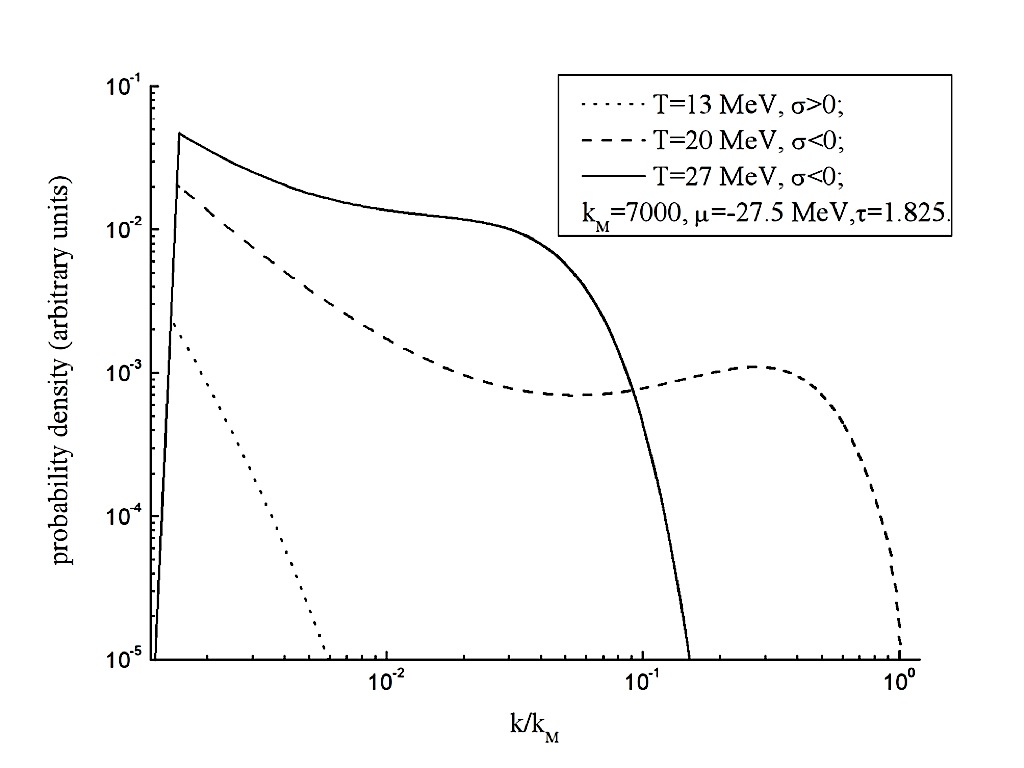} 
}  
 \caption{Fragment size distribution in the gaseous phase  is shown for a fixed baryonic chemical potential $\mu = -27.5$ MeV and three values of the  temperature $T$. The legend is similar to Fig. \ref{fig3}.  The  dotted  curve  
 in this figure corresponds to the solid curve in Fig. \ref{fig3}.  
}
  \label{fig4}
\end{figure}

\begin{figure}[t]
\centerline{
\includegraphics[height=11.11 cm]{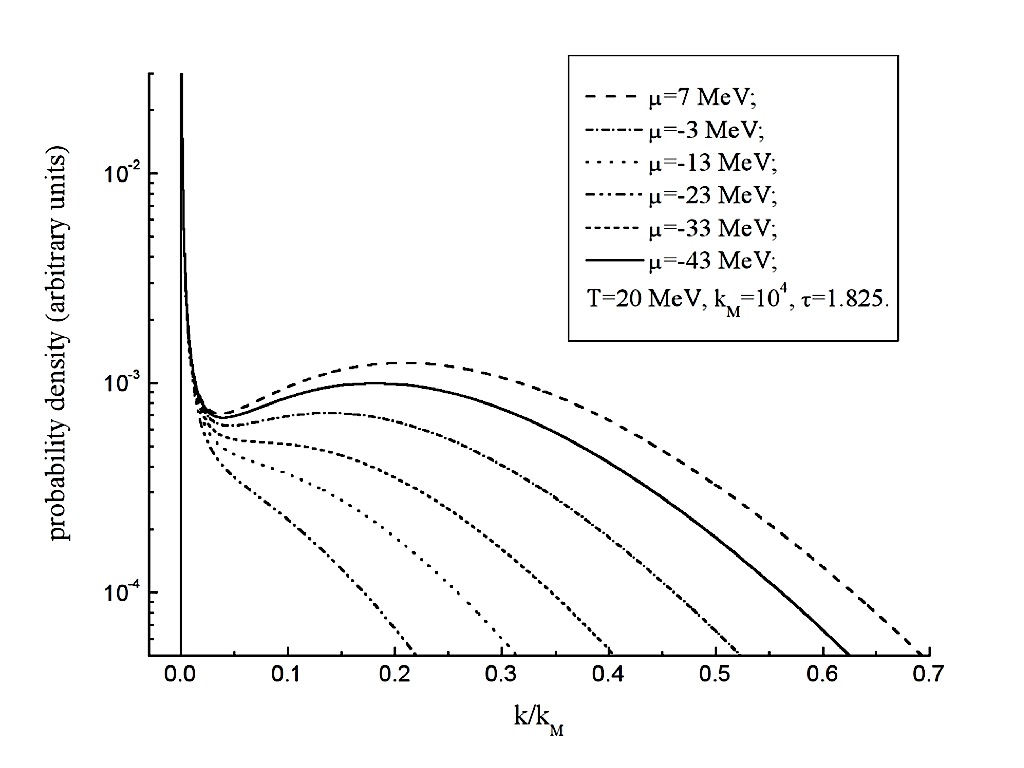} 
}  
 \caption{Fragment size distribution in the gaseous phase  is shown for a fixed temperature $T = 20$ MeV and several  values of the  baryonic chemical potential $\mu$.  The legend is similar to Fig. \ref{fig3}.  The principal difference with the distributions shown in Fig. \ref{fig3} is the presence of negative surface tension coefficient.  Note that the shown fragment size distributions  demonstrate a nonmonotonic dependence on the baryonic chemical potential.}
  \label{fig5}
\end{figure}

However, for the supercritical temperatures one finds the typical bimodal fragment distribution for a variety of temperatures and chemical potentials as one can see from Figs. \ref{fig4} and  \ref{fig5}.  It is necessary to stress that by construction at this  region the phase equilibrium is impossible due to negative surface tension coefficient, but the fragment distribution is bimodal and it very closely resembles the weighted  fragment size distributions found for the lattice gas model in \cite{Bmodal:Gulm07} shown there  in Fig. 5 and considered by the author of  \cite{Bmodal:Gulm07} as a clear PT signal  in a finite system. 
The bimodal distributions of the present model consist of three elements: there is a sharp peak at low $k$ values, then at intermediate  fragment sizes  there exists a local minimum, while at large   fragment sizes there is a wide maximum. A sharp peak reflects  a fast increase of the  probability density of dimers compared to the monomers (nucleons),  since the latter do not have the binding free energy and the surface free energy and, hence, the monomers are significantly suppressed in this region of thermodynamic parameters. On the other hand it is clear that the tail of fragment distributions in Figs. \ref{fig4} and  \ref{fig5} decreases due to the dominance of the bulk free energy and, hence, the whole structure at intermediate fragment sizes is due a competition between the surface free energy and two other contributions into the fragment free energy, i.e. the bulk one and the Fisher one.

Let us demonstrate now that the  bimodal 
  fragment size attenuation appears due to the negative value of the  surface tension coefficient, i.e. for $\sigma (T) < 0$.  In the latter  case the gaseous pressure exceeds that one of the liquid phase, i.e.  the effective chemical potential $\nu \equiv (p_l (T, \mu) - p_g (T, \mu) )\, b < 0$ is negative 
 \cite{SMM:Bugaev00,Bugaev:Thesis10}. Then 
  the unnormalized distribution of nuclear fragments 
with respect to the number of nucleons $k$ 
\begin{eqnarray}
\label{EqXI}
\omega (k) = \exp \left[  - \frac{|\nu|}{T} \, k + \frac{|\sigma|}{T} k^ \varsigma - \tau \ln k \, \right] \,,
\end{eqnarray}
has the local minimum at some value $k_{min}$ and the local maximum at  $k_{max} > k_{min}$. This can be shown by inspecting the logarithmic derivative of $\omega(k)$ with respect to $k$. Thus, the extremum condition for such a derivative 
gives us 
\begin{eqnarray}
\label{EqXIIa}
\left. \frac{\partial \ln \omega (k)}{\partial \, k}\right|_{k=k_E} =  - \frac{|\nu|}{T}   +  \frac{|\sigma|}{T} \, \frac{ \varsigma}{k^{1-\varsigma}_E}\,   - \frac{\tau}{k_E}  = 0 
\quad \Rightarrow \quad k_E =  \left[ \frac{\varsigma\,  |\sigma| }{|\nu| + \frac{\tau\, T}{k_E }}\right]^\frac{1}{1- \varsigma}
\,,
\end{eqnarray}
where the extremum is reached for $k = k_E$.  Let us show now that the expression for $k_E$ in (\ref{EqXIIa}) has two positive solutions. In the first case we assume that the Fisher term dominates over the bulk one, i.e.  $|\nu| \ll  \frac{\tau}{k_E }$, which may occur  only  for  small values of  $k_E$. Then neglecting the term  $|\nu|$ in the above expression  for  $k_E$ one finds 
\begin{eqnarray}
\label{EqXIIIa}
 k_{min} =  k_E \simeq   \left[ \frac{\tau \,T  }{\varsigma\,  |\sigma|}\right]^\frac{1}{\varsigma}
\,. 
\end{eqnarray}
The analysis of the second derivative of $\ln \omega(k)$ with respect to $k$ 
\begin{eqnarray}
\label{EqXIVa}
\left. \frac{\partial^2 \ln \omega (k)}{\partial \, k^2}\right|_{k=k_{min}} =  - \varsigma (1 -  \varsigma)\, \frac{ |\sigma|}{T\,k^{2-\varsigma}_{min}}\,   + \frac{\tau}{k^2_{min}}  =   \frac{ \varsigma \,\tau}{k^2_{min}} > 0 
\,,
\end{eqnarray}
shows that this derivative is always positive, i.e. there is a local minimum, for $\varsigma > 0$.  Note that Eq. 
(\ref{EqXIIIa})  allows one to roughly estimate  the   surface tension as   $\sigma \simeq  -  \frac{\tau\, T}{\varsigma\, k_{min}^\varsigma}$, if the  position of  the local minim  is known (for an  exact expression see below).

In the opposite case, if  the bulk free energy dominates over the Fisher term, i.e. for  $|\nu| \gg  \frac{\tau\, T}{k_E }$, which  occurs only  for  large values of  $k_E$, the solution for $k_E$
takes the form 
\begin{eqnarray}
\label{EqXVa}
 k_{max} =  k_E \simeq   \left[ \frac{\varsigma\,  |\sigma|  }{ |\nu|}\right]^\frac{1}{1-\varsigma}
\,, 
\end{eqnarray}
and,  therefore, the second derivative of $\ln \omega(k)$ with respect to $k$  can be written as 
\begin{eqnarray}
\label{EqXVIa}
\left. \frac{\partial^2 \ln \omega (k)}{\partial \, k^2}\right|_{k=k_{max}} =  - \varsigma (1 -  \varsigma)\, \frac{ |\sigma|}{T\, k^{2-\varsigma}_{max}}\,   + \frac{\tau}{k^2_{max}}  =  -  \frac{1}{k_{max}}  \left[ 
 \frac{(1 -  \varsigma)\,|\nu|}{T} -   \frac{\tau}{k_{max}}\right]
\,.
\end{eqnarray}
Now it is clear that the second derivative (\ref{EqXVIa}) is  negative for   $|\nu|(1 -  \varsigma) > \frac{\tau \,T}{k_{max}}$.  Note that the latter inequality cannot be fulfilled for $(1 -  \varsigma) \ll 1$ only, whereas for the typical 
SMM value $\varsigma \simeq \frac{2}{3}$ the inequality $|\nu|(1 -  \varsigma) > \frac{\tau \,T}{k_{max}}$ is obeyed due to 
adopted  assumption  $|\nu| \gg  \frac{\tau \,T}{k_{max} }$. Thus, at $k \simeq k_{max}$ the fragment distribution (\ref{EqXI}) has a local maximum. 
The existence of the distribution with the saddle like shape that has  both  a local minimum and  a local maximum which are clearly seen in Figs. \ref{fig4} and \ref{fig5}.

In fact, if the positions of  both local extrema are known, i.e. $k_{min}$ and  $k_{max}$  are known, for instance, from the experiment, then 
for a given temperature $T$ one can exactly  find  both $\nu$ and $\sigma$. To demonstrate this, we introduce a new variable $x$
\begin{eqnarray}
\label{EqXVIIa}
k_E^\varsigma  \equiv   \frac{\tau \, T}{\varsigma\, |\sigma|}(1 +x)  \,.
\end{eqnarray}
Then in terms of this variable the extremum condition (\ref{EqXIIa}) can written as 
\begin{eqnarray}
\label{EqXVIIIa}
 \frac{\tau \, T}{ |\nu|} x =  \left[  \frac{\tau \, T}{\varsigma\, |\sigma|}(1 +x) \right]^\frac{1}{\varsigma} \,.
\end{eqnarray}
since $k_E \equiv \frac{\tau \, T}{ |\nu|} x$. 
Denoting  the solutions of Eq. (\ref {EqXVIIIa}) as $x_1 = \frac {|\nu|}{\tau \, T}\, k_{min}$ and 
$x_2 = \frac{ |\nu|}{\tau \, T}\, k_{max} \equiv R\, x_1$ and dividing expression (\ref {EqXVIIIa}) for $x = x_2$ by the same expression for $x=x_1$, one obtains the following equation for $x=x_1$
\begin{eqnarray}
\label{EqXIXa}
 R =  \left[  \frac{1 + R \, x_1}{1 +x_1} \right]^\frac{1}{\varsigma} \quad \Rightarrow \quad x_1 = \frac{R^\varsigma-1}{R - R^\varsigma}\,, \quad x_2 = R\,\frac{R^\varsigma-1}{R - R^\varsigma} \,,
\end{eqnarray}
if the ratio $R \equiv \frac{x_2}{x_1} \equiv \frac{k_{max}}{k_{min}}$ is known from the fragment distribution. The above results allow us to explicitly  find  the effective chemical potential $\nu$ and the surface tension coefficient $\sigma$ as
\begin{eqnarray}
\label{EqXXa}
|\nu| =  \frac{\tau\, T}{k_{min}} \cdot \frac{R^\varsigma-1}{R - R^\varsigma}\,, \quad 
|\sigma| = \frac{\tau\, T}{\varsigma\, k_{min}^\varsigma} \cdot \left[ 1+ \frac {|\nu|}{\tau \, T}\, k_{min} \right]=   \frac{\tau\, T}{\varsigma\, k_{min}^\varsigma} \cdot \frac{R-1}{R - R^\varsigma}\,.  
\end{eqnarray}
These expressions can be useful for the   experimental data analysis. 

From the above analysis it is evident that the bimodal distributions demonstrated in Figs. \ref{fig4} and  \ref{fig5} have nothing to do with the PT existence, but appear due to the competition of the negative surface free energy with the positive free energy terms generated by the Fisher topological exponent and the bulk term, which, respectively, dominate at small and large values of fragment size.  Thus, we give an explicit example to the widely spread belief  \cite{Bmodal:Chomaz01,Bmodal:Chomaz03, Bmodal:Gulm04, Bmodal:Gulm07,THill:1} that a bimodal distribution of typical order parameter (size of fragment) is an exclusive signal of a first order  PT in finite systems.  Together with the authors of Refs.
\cite{Moretto:05, Bmodal:Anti05,FS:Bugaev07} we would like to stress that without studying  the nature 
of the bimodal distributions one  cannot claim that a PT is   its  only origin. 

Furthermore,  the existence of bimodal distributions without a PT completely breaks down the logic of T. Hill \cite{THill:1}. According to \cite{THill:1} the interface energy between two phases should essentially suppress the coexistence of two `pure'  phases, but the states at supercritical temperatures are, indeed, kind of the coexistence of two phases, but in an absence of a PT and, hence,   without an explicit surface separating  them.

\section{Bimodal distributions at finite volumes}

In this section we would like to thoroughly analyze the second typical mistake of the approaches \cite{Bmodal:Chomaz03,Bmodal:Gulm04, Bmodal:Gulm07,THill:1,DGross:1} based on bimodality properties of a first order PT in finite systems. In these approaches it is implicitly assumed 
that, like in infinite systems, in finite systems there exist exactly  two `pure' phases  and  they correspond to  two peaks in the bimodal distribution of the order parameter. The examples given in the preceding section correspond to the thermodynamic limit, although in actual simulations we used 
$7\cdot 10^3$ and $10^4$ particles.  We found that  further increase of  the size of the largest fragment $K(V)$ in 
(\ref{EqIII}) generates  the relative numerical  errors below $10^{-8}$ compared to the results obtained 
in the thermodynamic limit.  
In this section, however,   we consider smaller systems whose behavior is far from the thermodynamic limit.

In order to illustrate some  of the  results  which are necessary for a discussion of bimodality in finite systems we introduce the real $R_n$ and imaginary $I_n$ parts of  $\lambda _n = R_n + i I_n$ and  consider  Eq. (\ref{EqII})
as a system of coupled transcendental equations 
\begin{eqnarray}\label{EqXXIa}
&&\hspace*{-0.cm} R_n = ~ \sum\limits_{k=1}^{K(V) } \phi_k (T)
~ \exp \left[ \frac{Re( \nu_n)\,k}{T} \right]  \cos(I_n b k)\,,
\\
\label{EqXXIIa}
&&\hspace*{-0.cm} I_n = - \sum\limits_{k=1}^{K(V) } \phi_k (T)
~\exp \left[ \frac{Re( \nu_n)\,k}{T} \right]  \sin(I_n b k)\,,
\end{eqnarray}
where  for convenience we introduced   the following  set of  the effective chemical potentials  $\nu_n $ 
\begin{equation}\label{EqXXIIIa}
\nu_n  \equiv  \nu(\lambda_n ) = p_l(T,\mu) b  - (R_n + i I_n) b\,T  \,,
\end{equation}
and the reduced distribution for nucleons $\phi_1 (T) = \left(\frac{m T }{2 \pi} \right)^{\frac{3}{2} }  z_1 \exp((\mu -  p_l(T,\mu) b)/T)$. 

Consider the real root $(R_0 > 0, I_0 = 0)$, first. 
Similarly to the SMM \cite{SMM:Bugaev00},  for $I_n = I_0 = 0$ the real root $R_0$  of  the CSMM exists for any $T$ and $\mu$.
Comparison  of  $R_0$ from (\ref{EqXXIa}) with the expression for vapor pressure of the analytical SMM solution  \cite{SMM:Bugaev00}
indicates  that $T R_0$ is  a constrained grand canonical pressure of the mixture of ideal gases with the chemical potential $\nu_0$. 
Let us show that that  the gas singularity is always the rightmost one. First we assume that for the same set of $T, \mu$ and $V$ there exists a complex root $R_{n>0}$ which is the rightmost one compared to $ R_0$, i.e. 
$R_{n>0} > R_0$ for  $I_{n>0} \neq 0$. Then one immediately concludes that  $Re(\nu_{n>0}) < Re(\nu_{0})$, but in this case
for  $n>0$  one obtains 
\begin{equation}\label{EqXXIVb}
R_n = \sum\limits_{k=1}^{K(V) }\phi_k (T)\,
{\textstyle \exp \left[ \frac{Re(\nu_{n})\, k}{T} \right] \, \cos(I_n b k) } < \sum\limits_{k=1}^{K(V) }\phi_k (T) \,
{\textstyle \exp \left[  \frac{Re(\nu_{0})\, k}{T} \right] }= R_0 \,,
\end{equation}
i.e. we arrive at a contradiction with the original assumption. 

Note, however,  that assuming an opposite inequality  $R_{n>0} <  R_0$ for  $I_{n>0} \neq 0$ and  $I_{0} = 0$, one cannot get a contradiction,  since a counterpart of  the  inequality  (\ref{EqXXIVb}) cannot be established  for  $Re(\nu_{n>0}) >  Re(\nu_{0})$ due to the fact that 
for $I_{n>0} \neq 0$ some of  the $k$-values  in the sum in   Eq. (\ref{EqXXIIa}) unavoidably 
would generate  the inequality $\cos(I_n b k) < 1$. This means that the gas singularity is always the rightmost one. Such a fact plays a decisive role in a formulating the finite volume analogs of phases  
\cite{Bugaev:CSMM05}
and it will be exploited  below as well. 

Since Eq.  (\ref{EqXXIIa}) is not changed under the substitution $I_n \leftrightarrow - I_n$ then the complex roots of the system 
(\ref{EqXXIa}), (\ref{EqXXIIa}) are coming in pairs only. This is an evident consequence of the fact that the grand canonical partition 
(\ref{EqI}) must be real. 
Now it is also apparent  that all the roots can be classified  according to  a descending order of their real parts. 

A rigorous mathematical  scheme to identify the analogs of phases in finite systems for the partitions (\ref{EqI})-(\ref{EqIV})   was worked out in 
\cite{Bugaev:CSMM05, FS:Bugaev07,Bugaev:Thesis10}. It is based on the number of roots of  the system  
 (\ref{EqXXIa}),  (\ref{EqXXIIa}) for a given set of grand canonical variables $T, \mu$ and $V$. Thus,  a single real solution $\lambda_0 = R_0$ with $I_0=0$ of the system  
 (\ref{EqXXIa}),  (\ref{EqXXIIa})
corresponds to a gaseous phase, since its pressure, indeed, looks like a pressure of a mixture of ideal gases with a single value of  the effective baryonic chemical potential $\nu_0$ defined by (\ref{EqXXIIIa}). 
If the system  
 (\ref{EqXXIa}),  (\ref{EqXXIIa})  has one real solution $\lambda_0$ and any natural  number  $ n=1, 2, 3, ...$ of  complex conjugate pairs of roots $\lambda_{n\ge 1}$ then the  corresponding partition (\ref{EqI})  describes  a mixture of a gaseous phase with a set of  metastable states which are not in a true chemical equilibrium with the gas, since the real parts of their  free energy  $ -T V R_{n> 0}$  are larger than the corresponding value for the gaseous phase, i.e. 
$ -T V R_{n>0} > -T V R_0$. The absence of  a true chemical equilibrium between these 
metastable states
and  the gas is also seen from that the fact 
the real parts of  their   effective chemical potential $\nu_n $   of 
 is larger than the value of the  effective chemical potential  of the gaseous phase $\nu_0$, i.e. $Re(\nu_{n>0}) > \nu_0$.
 A finite system analog of a  fluid phase  corresponds to an infinite number of the complex roots of the system  
 (\ref{EqXXIa}),  (\ref{EqXXIIa}), but it exists at infinite pressure only. 

\begin{figure}[t]
%
%
\centerline{
\includegraphics[height=10.10 cm]{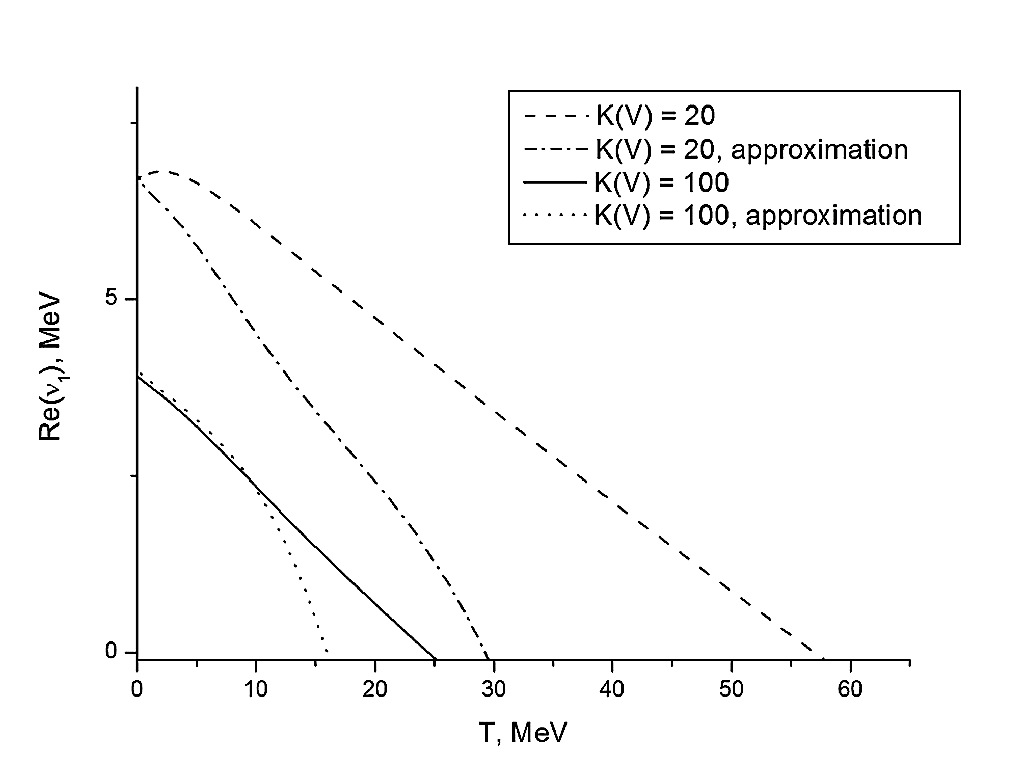} 
}  
 \caption{
The finite volume analog of the phase diagram in $T-Re(\nu_1)$ plane for given values of $K(V) = 20$ (dashed curve)
and $K(V) = 100$ (solid curve).
Below  each of these phase boundaries there exists a gaseous phase only, but at and above each curve there
are  three or more solutions of the system (\ref{EqXXIa}),  (\ref{EqXXIIa}). These solutions describe the 
states that  can be identified as  a finite volume analog of a  mixed phase.  
The additional curves correspond to the approximation (\ref{EqXXIXa}).
}
  \label{fig6}
\end{figure}

\begin{figure}[t]
%
%
\centerline{
\includegraphics[height=10.10 cm]{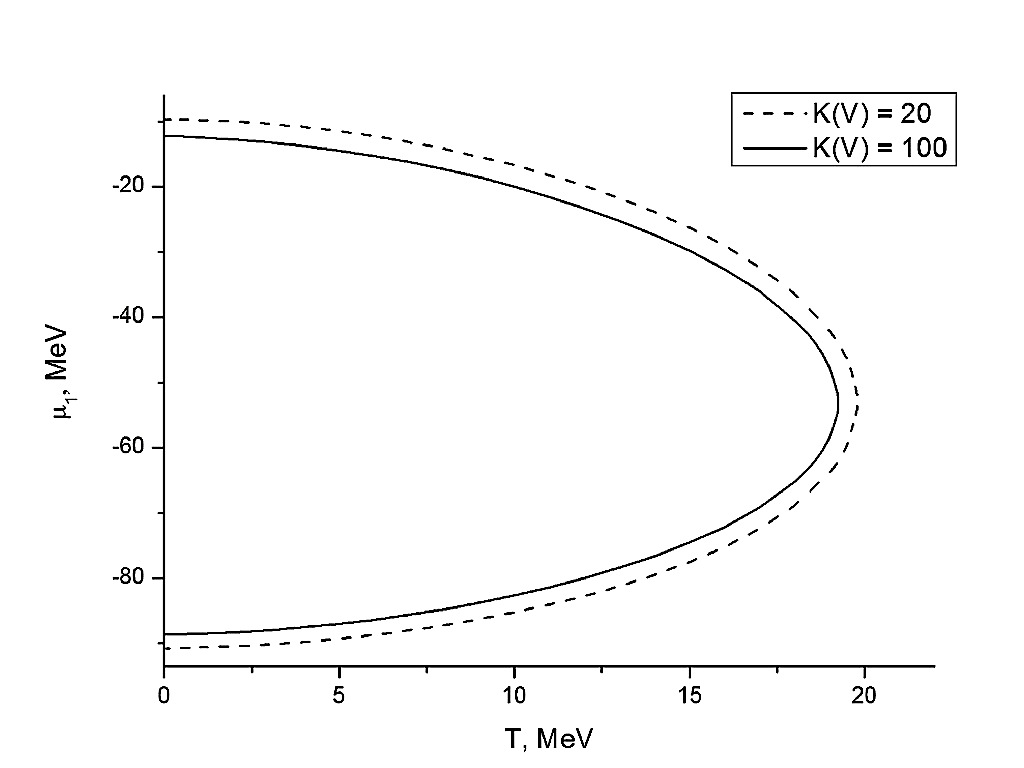} 
}  
 \caption{
The image of  the finite volume analog of the phase diagram  $T-Re(\nu_1)$  of  Fig. \ref{fig6} is shown in terms of the usual variables  $T$ and $\mu$. Note that for finite  $K(V)$ the solutions $\mu_1(T)$ do not exist for some temperatures   $\max(T) > T_{cep}$ and, thus, the  both phase equilibrium curves 
of Fig. \ref{fig1} form a continuous phase  diagram   for a finite system. 
}
  \label{fig7}
\end{figure}

Using this   scheme, one can build up the finite system analog of the $T-\mu$ phase diagram. Indeed,  the curve $Re(\nu_1(T))$ 
divides the  temperature-chemical potential plane  into three regions: for the region $Re(\nu_n) < Re(\nu_1(T))$ there is only a single solution of the system  (\ref{EqXXIa}),  (\ref{EqXXIIa}) which describes the gaseous phase, at the curve $Re(\nu_n) = Re(\nu_1(T))$ 
there are exactly three roots of the system (\ref{EqXXIa}),  (\ref{EqXXIIa}) while 
 above for  $Re(\nu_n) >  Re(\nu_1(T))$ there are five or more roots of this system, which corresponds to a finite volume analog of mixed phase. Fig. \ref{fig6} shows such a curve $Re(\nu_1(T))$. The principal difference with the thermodynamic  limit discussed in the preceding section  is that  for finite volumes the effective chemical potential in the gaseous phase can be positive, i.e. for some temperatures one has $\nu_0 > 0$. 
Knowing the values of  $Re(\nu_1(T))$ and $R_1(T)$, one can find the corresponding value of the  liquid pressure, which, in its turn,  allows one to determine the curve $\mu_1 (T)$  from the liquid phase equation of state (\ref{EqV}). 
 Such  curves  $\mu_1 (T)$ are shown in Fig. \ref{fig7} for  two values  of the maximal fragment size $K(V)$. Comparing the $T-\mu$ phase diagrams of  Fig. \ref{fig7} with that ones shown in Fig. \ref{fig1}, one can see that   for temperatures below $T_{cep}$ all  the curves  are  quantitatively  similar to each other 
 even for  a small system with $K(V) = 20$. However,  in contrast to the thermodynamic limit phase diagram of  Fig. \ref{fig1},  for  considered finite systems the curves $\mu_1(T)$ for the nuclear matter 
 and `antinuclear'  matter are connected with each other at temperatures about $T_{cep}$.

 It is necessary to stress that, in contrast to the infinite systems,   the partial pressures  $T  R_n$ of the states
 $ n=0, 1, 2, 3,..$ that belong to the same grand canonical partition of a  finite system (\ref{EqI}) do not coincide with each other and, therefore,  in contrast to the beliefs of the authors of  \cite{Bmodal:Chomaz03,Bmodal:Gulm04, Bmodal:Gulm07,THill:1},  the statistical weights of  the gaseous phase $(n=0)$  and the states with  $n \ge 1$ can be quite different. Moreover, although the state with $n=0$ is a gaseous phase,  the states  with  $n \ge 1$  cannot be  identified  as  a `pure' liquid,  since they have different 
partial pressures  and different decay/formation times defined via the imaginary part of the free energy as $\tau_n   \equiv \left[ I_n b T \right]^{-1}$ \cite{FS:Bugaev07, Bugaev:CSMM05,Bugaev:Thesis10}.
Furthermore, in finite systems  even the gaseous phase differs from that one existing in the thermodynamic limit,  
since,  as one can see from Fig. \ref{fig6},  for  finite volumes $V$ the effective chemical potential can be positive, i.e. $\nu_0 > 0$, and this case corresponds  to entirely different  distribution of fragments.

\begin{figure}[t]
%
%
\centerline{
\includegraphics[height=10.10 cm]{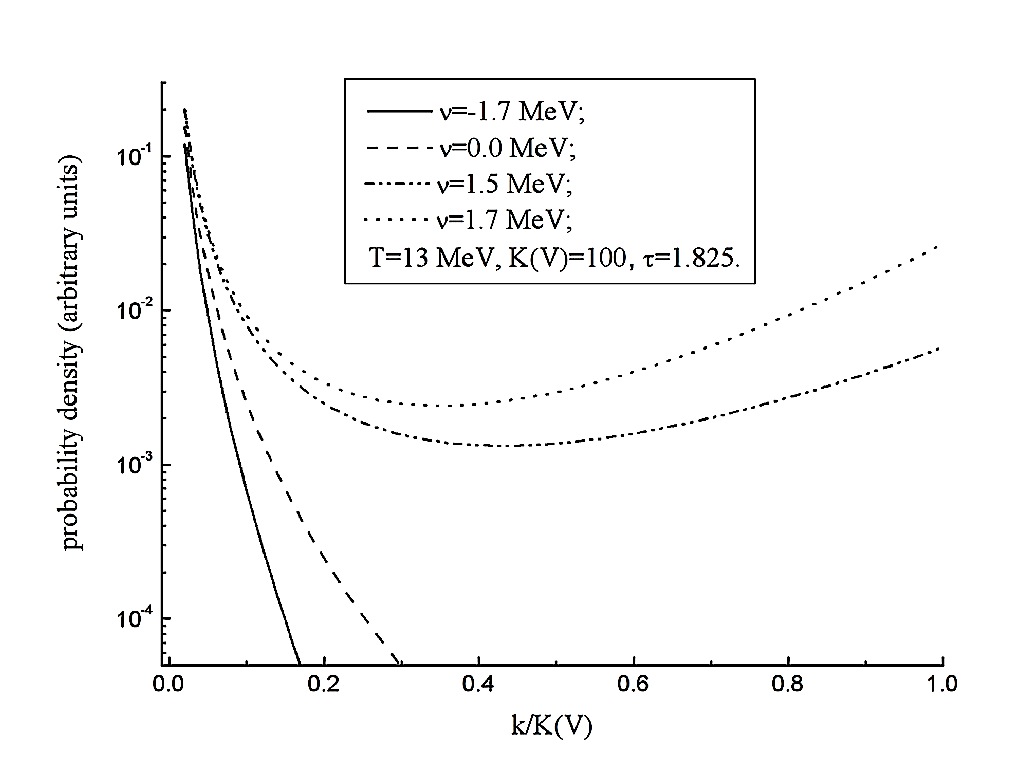} 
}  
 \caption{
Typical fragment size distributions existing in  a finite analog of  gaseous phase  are shown for a fixed temperature $T = 13$ MeV and different value of the effective  chemical potential $\nu_0$.
For positive values of  $\nu_0$ the fragment distribution has a bimodal like shape, although it is still a gas of all fragments.
The maximal size of nuclear fragment   is $K(V) = k_M =100$ nucleons.}
  \label{fig8}
\end{figure}

Indeed, as one can see from Fig. \ref{fig8} for  positive values of  the effective  chemical potential $\nu_0$ the fragment size distributions in a finite analog of  gaseous phase acquires a bimodal like shape without any PT. 
Existence of such distributions is another explicit counterexample against Hill belief  \cite{THill:1} that the
bimodal distributions can be used to unambiguously characterize a PT in finite systems.

Since an existence of the states with  $\nu_0 > 0$   is of principal importance for this study, here we would like to demonstrate this fact analytically. 
For this purpose we
consider  the limit  $Re(\nu_n)  \gg T $ for all  $0 < n < N 0$ with $N \gg1$.  
For instance, this is a typical situation for low temperatures $T$ or it can appear at high baryonic densities 
existing inside of  a mixed phase.  
It is clear, that in this limit
the leading contribution to the right hand side  of (\ref{EqXXIIa}) corresponds to the harmonic with $k = K (V)$,
and, consequently,  an exponentially large amplitude of this term
can be only  compensated by  a vanishing value of  $\sin\left( I_n \, b K(V)  \right)$,  
i.e.   $I_n \, b K  =  \pi n + \delta_n$
with $|\delta_n| \ll \pi$ (hereafter we will analyze only 
the branch $I_n > 0$).  
Keeping the leading term on the right hand side of  (\ref{EqXXIIa}) and solving for $\delta_n$, one finds  \cite{Bugaev:CSMM05,Bugaev:Thesis10,Bugaev:Nucleation11}   
\begin{eqnarray}\label{EqXXVb}
\hspace*{-0.0cm}
I_n & \approx & \frac{2 \pi\, n + \delta_n}{K(V)\,b} \approx  \frac{2 \pi\, n }{ K(V) \, b}\left[1 - \frac{1 }{K(V)\,b\, R_n}  \right] \,, \\
\label{EqXXVIb}
%
\delta_n &\approx  & -  \frac{ 2  \pi n }{ K(V)\, b\, R_n}  \,,    \\
\label{EqXXVIIb}
R_n & \approx &   \phi_K (T)  ~ \exp \left[ \frac{Re(\nu_n )\,K(V)}{T} \right]   \,,
\end{eqnarray}
where the results are given for the branch of positive $R_n$ values. 

Since for large volumes $V$  the  negative values of $R_n$ cannot contribute to the 
grand canonical partition (\ref{EqI}), here we  analyze only  values of $n$ which generate  $R_n > 0$.
In this case substituting the reduced distribution (\ref{EqIX})  into Eq.  (\ref{EqXXVIIb}) one obtains the leading terms for the  partial pressure of $n$-th state 
\begin{eqnarray}
T R_n & \approx &  p_l(T,\mu)  -  \frac{ T}{ b\, K(V) } 
 \ln \left|  \frac{ R_n }{  \phi_K (T)     }  \right|  \nonumber \\
 & \approx &  p_l(T,\mu) - \frac{\sigma(T)}{ b\, [K(V)]^{1- \varsigma}} - T \left[ \frac{\ln |
 \left(\frac{2 \pi}{ m T } \right)^{\frac{3}{2}} R_n  | + \tau \ln K(V)}{b\, K(V) } \right] \,, 
  \label{EqXXVIIIa}
\end{eqnarray}
under the inequalities  $ Re( \nu ) \gg T $ and $K(V) \gg 1$.  This equation clearly shows that for $K(V) \gg1 $ and $\varsigma = \frac{2}{3}$ the 
$n$-th state corresponds to  a finite droplet of a radius of $K(V)^\frac{1}{3}$ nucleon radii having    a  volume pressure of an infinite liquid droplet  which is 
corrected by the Laplace surface pressure (the second term on the right hand side of  (\ref{EqXXVIIIa})). 
In fact, such states correspond to a mixed phase dominated by a heaviest fragment. 
This is clearly seen from (\ref{EqXXVIIIa})  at low temperatures. Indeed, for $T\rightarrow 0$ the left hand side 
of  (\ref{EqXXVIIIa}) and the last term on the right hand side of it vanish and we obtain that equations for all $R_{n>0}$ degenerate into the same expression $p_l(0,\mu_1)  - \frac{\sigma(0)}{ b\, [K(V)]^{1- \varsigma}} \approx 0$, which is a condition of vanishing total pressure of the finite liquid drop, 
where the chemical potential $\mu_1$ corresponds to $R_1$.  
 In other words, the vanishing total pressure of  the $n$-th state is the   mechanical stability condition of   mixed phase, since at $T\rightarrow 0$  the gaseous phase pressure is zero. A few examples of $\mu_1 (T)$ are depicted in Fig. \ref{fig7}.

Also Eq. (\ref{EqXXVIIIa})  
obviously  demonstrates that in the thermodynamic limit $K(V) \rightarrow \infty$ an infinite number of metastable states with partial pressures $T R_{n>0}\rightarrow p_l(T,\mu)$ go to the real axis of the complex $\lambda$-plane,  since in this limit  $I_{n>0}\rightarrow 0$ in (\ref{EqXXVb}),  and, hence, they  form a pole of infinite order at 
$\lambda_{n>0} = p_l(T,\mu)/T$, i.e. they form an essential singularity of the isobaric partition function 
\cite{FS:Bugaev07, Bugaev:CSMM05,Bugaev:Thesis10,Bugaev:Nucleation11} which, in contrast to a simple pole of a gaseous phase $\lambda_{0} =R_0$,  describes a liquid phase.  

From 
Eq.  (\ref{EqXXVIIIa})   one can  get  the effective chemical potentials $Re( \nu_{n>0} )$ of these $n$-states as
\begin{eqnarray}
Re( \nu_{n>0}  ) \approx  \frac{\sigma (T)}{[K(V)]^{1- \varsigma}  } + T \left[ \frac{\ln |
 \left(\frac{2 \pi}{ m T } \right)^{\frac{3}{2}} R_{n>0}  | + \tau \ln K(V)}{K(V) } \right] \, ,
  \label{EqXXIXa}
\end{eqnarray}
from which one can immediately deduce  that for low temperatures and for  $K(V) \gg1 $ the real part of 
$\nu_{n>0} $
is solely defined by the sign of the surface tension coefficient, i.e. from $\sigma (T) > 0$ it follows that  $Re( \nu_{n>0}  ) > 0$.  In the thermodynamic limit $K(V) \rightarrow \infty$ Eq. (\ref{EqXXIXa}) recovers the usual SMM result that the effective  chemical potential vanishes only  at the phase equilibrium line \cite{SMM:Bugaev00}.

Furthermore,  in the limit $T \rightarrow 0$ from (\ref{EqXXIXa}) one finds that 
\begin{eqnarray}
Re( \nu_{1}  ) \approx Re( \nu_{2}  ) \approx Re( \nu_{3}  )  \approx ... \approx  Re( \nu_{n}  )  \approx b \, p_l(0,\mu_1)  \approx \frac{\sigma (0)}{[K(V)]^{1- \varsigma}  }  \, ,
  \label{EqXXX}
\end{eqnarray}
i.e. the real parts  of all  effective chemical potential states  are tending to match at  vanishing temperatures independently on the values of $R_{n>0}$ for  $\mu = \mu_1$ introduced earlier. From (\ref{EqXXX}) one can easily show that for $ \nu_{0} < Re( \nu_{1}  ) $ the liquid droplet cannot exist in the limit $T \rightarrow 0$. Suppose, on the  contrary, that this is possible. 
Then such a situation  can occur only for  some chemical potential $\mu^\prime$  defined as $ \nu_{0} = b \, p_l(0,\mu^\prime)  $. Obviously $\mu^\prime < \mu_1$, since  for the 
equation of state of liquid (\ref{EqV})  its pressure $p_l(0,\mu) $ is a monotonically increasing function of  chemical potential  $\mu$.
However, as we showed above the total pressure of such  finite droplet is $p_l(0,\mu^\prime)  - \frac{\sigma(0)}{ b\, [K(V)]^{1- \varsigma}}< 0$ and, hence, such a droplet is mechanically unstable and it cannot exist
under such conditions.
 On the other hand, for $\mu^\prime > \mu_1$ or, equivalently,  for $ \nu_{0} = b \, p_l(0,\mu^\prime)  >  Re( \nu_{n>0}  )  \approx b \, p_l(0,\mu_1)  $ the solution $R_0$  always exists which means that the finite volume analog of  the gaseous phase exists together with the solutions $R_{n>0}$ describing the finite droplet.  These  are simple    physical  arguments   that  $Re( \nu_{1}(T)) $ is a finite volume analog of the $T-\mu$ diagram of  the first order PT at $T\rightarrow 0$.  More formal arguments can be found in \cite{FS:Bugaev07, Bugaev:CSMM05,Bugaev:Thesis10}.

As one can see from Fig. \ref{fig6} the expression (\ref{EqXXIXa}) approximately  reproduces  the numerical solution  of the system (\ref{EqXXIa}),  (\ref{EqXXIIa}) for $Re(\nu_1)$.  Moreover, this figure clearly demonstrates that at low temperatures  the  condition $Re(\nu_1) \gg T$ is obeyed and, hence,  the approximation  (\ref{EqXXIXa}) works   well even for  a small system with  $K(V) = 20$.  For a larger system with $K(V) = 100$ Eq. (\ref{EqXXIXa})   
correctly reproduces  the temperature dependence of  $Re(\nu_1(T)) $  for all temperatures below 12 MeV, although in this case the inequality $Re(\nu_1(T))  \gg T$ is not  obeyed. 

\begin{figure}[t]
%
%
\centerline{
\includegraphics[height=10.10 cm]{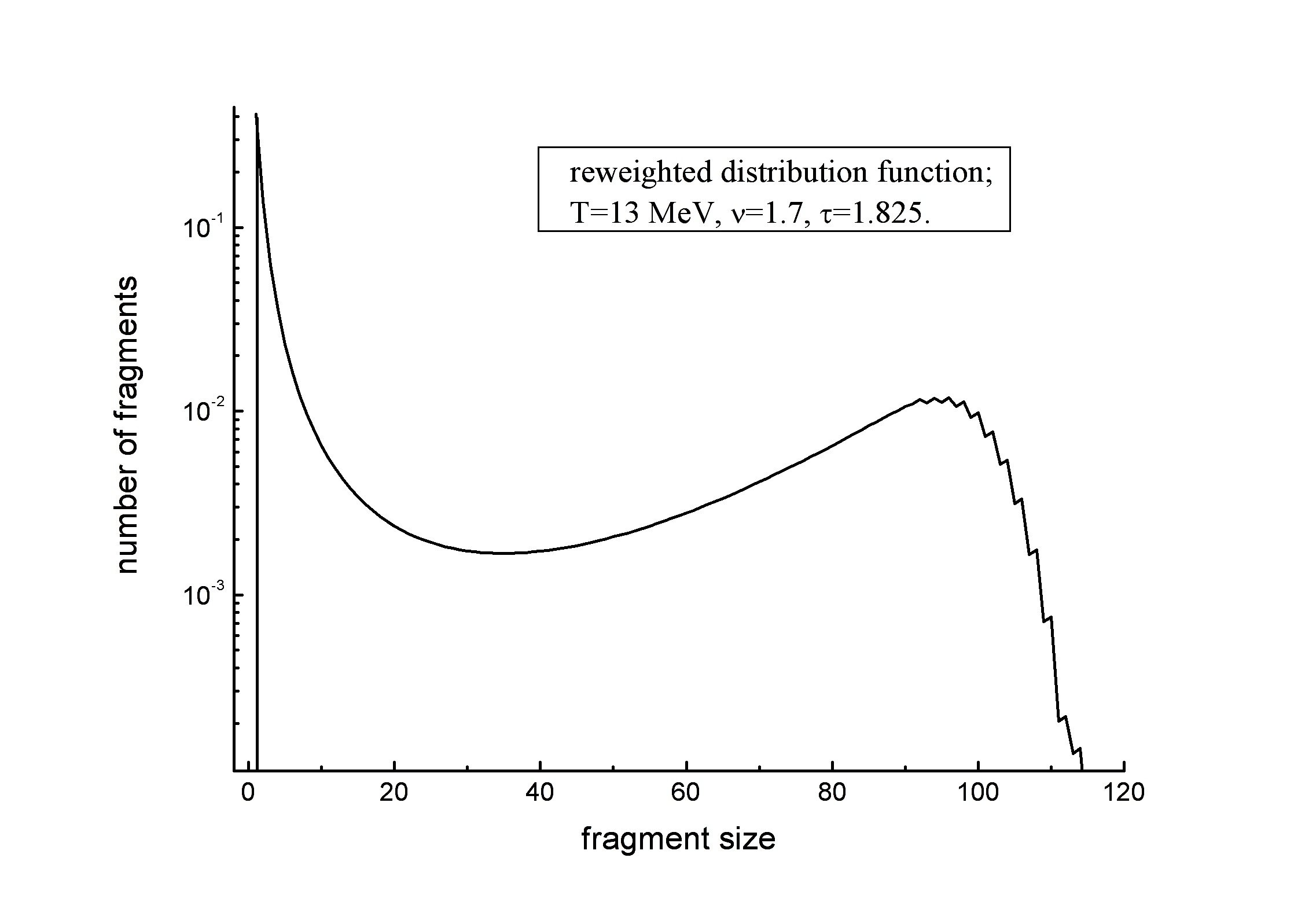} 
}  
 \caption{
The reweighted fragment size distribution for  a finite analog of  gaseous phase.  The original 
fragment size distribution corresponds to the parameters $T=13 MeV$ and $\nu_0 = 1.7$ MeV 
(see the corresponding curve in Fig. \ref{fig8}) , but for $K(V) \in [ 85; 115]$  values distributed normally
 with the mean value $\bar K(V) = 100$ and a dispersion $5$. 
}
  \label{fig9}
\end{figure}

Also the above analysis demonstrates that the finite volume analog of the tricritical point with the parameters $Re(\nu) = 0$ and $\sigma(T) =0$, i.e. a state at which 
the gaseous phase pressure  coincides with the pressure of infinite liquid droplet and the surface free energy  is zero, belongs to a finite volume analog of a gaseous  phase, since according to the above analysis 
such equalities for finite systems can be achieved   only at $T=T_{cep}$ and only for  $\nu_0 = Re(\nu_0) =  0$.  Note that at the finite volume analog of the tricritical point  the size distribution of the fragments is purely 
power like. It is hoped that such a feature can be helpful for an  experimental identification of the tricritical point  in the experiments.

An  existence of  the  gaseous  phase  with  $\nu_0 > 0$ in finite systems clearly indicates  the principal difference between the properties of gaseous phases existing in finite  and in infinite volumes. And this principal  difference can be seen in the fragment distributions shown in Figs. \ref{fig3} and \ref{fig8}. Indeed, the fragment size 
distributions depicted in Fig.  \ref{fig3} are monotonically decreasing ones, even taken at the boundary between the macroscopic gaseous phase and macroscopic mixed phase, whereas
for $\nu_0 > 0$   the fragment size  distributions of  Fig.  \ref{fig8} have a bimodal shape.  The latter 
might not look as a canonical bimodal shape, but if one accounts for the fluctuation of the maximal number 
of nucleons in the system which is similar to the number of participating nucleons in the nuclear reaction, 
then the resulting distribution may look much more similar to those one discussed in Refs. \cite{Bmodal:Gulm04,Bmodal:Gulm07,Bmodal:Lopez06}. In  Fig. \ref{fig9} we show such a reweighted distribution which was constructed from fifteen distributions having the same values of $T= 13$ MeV and 
$\nu_0 = 1.7$ MeV, but for  the parameter $K(V) $ distributed normally  in the range $K(V) \in [ 85; 115]$ 
 with the mean value $\bar K(V) = 100$ and a dispersion $5$.  Such a reweighting models  the 
possible dynamical fluctuations  of the impact parameter in the nuclear reaction.  The example of 
Fig. \ref{fig9} demonstrates that  the observed fragment size distribution does differ from the original statistical  fragment size distribution due to weak  dynamical fluctuations of  the impact parameter. 
The effect of the dynamical fluctuations of the initial temperature (which appears at the moment of thermal equilibrium)  that is well-known in the high energy  hadron and nuclear  collisions  \cite{Wilk:2008} can be even more dramatic and it can essentially  modify the original statistical  fragment size distribution. The worst is that  it is entirely unclear how  this cause or/and the other possible physical ones   like a collective flow and   its  instabilities modify   the original statistical  fragment size distribution before it is measured by a detector. 
Therefore, from this example and the counterexamples given  above  we conclude that  it is hard to believe  that the  theoretical schemes suggested in  \cite{Bmodal:Chomaz01,Bmodal:Chomaz03, Bmodal:Gulm04, Bmodal:Gulm07}  to manipulate  with the observed  data  are, indeed, able to elucidate any essential PT related characteristics  of the statistical distributions from the measured  data.

\section{Conclusions}\label{secConclusions} 

In the present work we gave two explicit counterexamples to the widely spread beliefs 
\cite{Bmodal:Chomaz01,Bmodal:Chomaz03, Bmodal:Gulm07,Bmodal:Lopez06, THill:1} 
about  an exclusive role of bimodality as the first order PT signal 
and showed  that the bimodal distributions can naturally  appear both  in  infinite and in finite systems without a PT. In the first  counterexample 
a bimodal distribution is generated at the supercritical temperatures by the negative values of the surface tension coefficient. This result is in line with the previously discussed role of  the competition between the 
volume and the surface parts of  the system free energy \cite{Moretto:05,FS:Bugaev07}.
In the second considered counterexample a bimodal fragment distribution is generated by positive values of the effective chemical potential in a  finite volume analog of  a gaseous phase. The latter was provided by an exact analytical solution of the CSMM for finite systems \cite{Bugaev:CSMM05,FS:Bugaev07} which was successfully generalized here for more realistic equations of state of  the compressible nuclear liquid  and for more realistic treatment of the surface tension free energy. 

Also here we gave  analytic results  showing for the first time that for finite, but large  systems,  the value
of the effective chemical potential on the finite volume analog of the $T-\nu$ phase diagram  \cite{Bugaev:CSMM05,FS:Bugaev07} is solely defined by the surface tension coefficient and by the radius of the largest fragment. The derived  analytical formulas for partial pressures of the metastable states belonging to the same grand canonical partition give an explicit example that, on the contrary to the beliefs of  Refs.  
\cite{Bmodal:Chomaz01,Bmodal:Chomaz03, Bmodal:Gulm04, Bmodal:Gulm07, THill:1}, in finite systems there are no two `pure' phases as it is the case in the thermodynamic limit. At finite pressures  the liquid-like finite droplet appears only as a part of a finite volume analog of a mixed phase. Additionally,  here we demonstrated that for positive values of the effective chemical potential $\nu_0$ the properties of the gaseous phase in finite systems drastically differ from its properties in the thermodynamic limit. 
The bimodal  fragment size distributions depicted in Figs. \ref{fig8} and \ref{fig9}  cannot exist in the gaseous phase treated  in the thermodynamic limit (see Fig.  \ref{fig3} for comparison).

The above results are in line with the critique  \cite{Moretto:05, Bmodal:Anti05,FS:Bugaev07} of  a bimodality as a reliable  signal of the PT existence in finite systems. 
Once more we have to stress that without studying  the nature 
of the bimodal distributions one  cannot claim that a PT is   its  only origin.
An interesting result on the bimodality absence in the systems indicating a possible PT existence  in multifragment production in heavy-ion nuclear collisions  was reported in \cite{Frankland:2004}. This is  an additional counterexample to the widely spread belief  on an exclusive role of bimodality as a PT signal in finite systems.

Therefore,
all the counterexamples obtained  in this work on the basis of an exactly solvable statistical model known as the CSMM allow us to  conclude that  it is rather  doubtful  that the  theoretical schemes  invented in Refs. \cite{Bmodal:Chomaz01,Bmodal:Chomaz03, Bmodal:Gulm04, Bmodal:Gulm07}  to manipulate  with the observed  data    are, indeed, able to elucidate the reliable  PT  signals  from the measured  data.


\vspace*{4mm}

{\bf Acknowledgments.} 
We  appreciate the valuable comments of I. N. Mishustin and    L. M. Satarov. 
Also the authors   acknowledge  a partial  support of the Program ``Fundamental Properties of Physical Systems 
under Extreme Conditions'' launched by  the Section of Physics and Astronomy  of
the National Academy of Sciences of Ukraine.



\begin{thebibliography}{99}

\bibitem{Bmodal:Chomaz01}
%
Ph. Chomaz, F. Gulminelli and  V. Duflot,  Phys. Rev. E {\bf 64},   046114 (2001).

\bibitem{Bmodal:Chomaz03}
%
Ph. Chomaz and  F. Gulminelli,
 Preprint GANIL-02-19 (2002).

\bibitem{Bmodal:Gulm04}
%
F. Gulminelli, Ann. Phys. Fr. {\bf 29}, 6 (2004) and references therein.

\bibitem{Bmodal:Gulm07}
%
F. Gulminelli,
 Nucl. Phys. A {\bf 791},  165  (2007).

\bibitem{Bmodal:Lopez06}
%
O. Lopez and  M. F. Rivet,
Eur. Phys. J. A {\bf 30}, 263 (2006)  and references therein. 

\bibitem{Bmodal:Indra06}
%
M. Pichon et al. (INDRA and ALADIN Collaborations), 
Nucl. Phys. A {\bf 779}, 267 (2006).

\bibitem{Bmodal:Bruno08}
%
M. Bruno et al., Nucl. Phys. A {\bf 807}, 48 (2008).

\bibitem{Bmodal:Indra09}
%
E. Bonnet et al. (INDRA and ALADIN Collaborations), 
Phys. Rev. Lett. {\bf 103},  072701  (2009).  



\bibitem{YangLee:52}
%
C. N. Yang and T. D. Lee, Phys. Rev. {\bf 87}, 404 (1952). 

\bibitem{Moretto:05}
%
L. G. Moretto, J. B. Elliott and L. W. Phair,
{ Mesoscopy and Thermodynamics},  proceedings  of  the conference 
{\it ``World Consensus Initiative III''}, Texas A \& M University, College Station,
Texas, USA, February 11-17, 2005
({see  http://cyclotron.tamu.edu/wci3/newer/chapVI\_4.pdf}).



\bibitem{Bmodal:Anti05}
%
O. Lopez, D. Lacroix, and E. Vient,
Phys. Rev. Lett. {\bf 95}, 242701 (2005).  

\bibitem{FS:Bugaev07}
%
K. A. Bugaev, 
Phys. Part. Nucl. {\bf 38},  447  (2007).



\bibitem{THill:1}
%
T. L. Hill,  {\it Thermodynamics of small  systems} (Dover, New York 1994).

\bibitem{DGross:1}
%
D. H. E. Gross, {\it Microcanonical Thermodynamics: Phase Transitions in Finite Systems}, Lecture Notes in Physics,  vol. 66 (World Scientific, 2001). 

\bibitem{Bugaev:CSMM05}
%
K. A. Bugaev,
Acta. Phys. Polon. B {\bf  36}, 3083 (2005).

\bibitem{SMM:Bondorf95}
%
J. P. Bondorf et al., Phys. Rep.  {\bf 257}, 131 (1995).

\bibitem{SMM:Bugaev00}
%
K. A. Bugaev,
M. I. Gorenstein, I. N. Mishustin and W. Greiner,
Phys. Rev. C {\bf 62},  044320 (2000);
arXiv:nucl-th/0007062 (2000);
%
Phys. Lett. B {\bf  498},  144 (2001);
arXiv:nucl-th/0103075 (2001).

\bibitem{Reuter:01}
%
P. T. Reuter and K. A. Bugaev,
Phys.\ Lett.\ B {\bf 517},  233 (2001).

\bibitem{HDM:Bugaev05} 
K. A. Bugaev, L. Phair and J. B. Elliott,
  Phys.\ Rev.\ E {\bf 72}, 047106 (2005).
  
   
\bibitem{HDM:Bugaev07} 
K. A. Bugaev  and J. B. Elliott,
Ukr. J. Phys. {\bf 52},  301 (2007).

\bibitem{Bugaev:Thesis10}
K.~A.~Bugaev,
arXiv:1012.3400 [nucl-th].


\bibitem{Bugaev:Nucleation11}
%
K. A. Bugaev, A. I. Ivanitskii, E. G. Nikonov,  A. S. Sorin and G. M. Zinovjev,
Can We Rigorously Define Phases in a Finite System?, 
Chapter 18 of the Proceedings of the XV-th Research Workshop 
{\it ``Nucleation Theory and Applications"}, held at  JINR,  Dubna, Russia, April 1- 30, 2011, 
edited by J. W. P. Schmelzer, G. Ropke, V. B.  Priezzhev, Dubna JINR, 2011;
 arXiv:1106.5939 [nucl-th] 

\bibitem{SMM:simple98}
%
S. Das Gupta and A.Z. Mekjian, Phys. Rev.  C {\bf 57}, 1361 (1998).

\bibitem{ISIS:99}
%
L. Beaulieu et al. (ISiS Collaboration), Phys. Lett. B {\bf 463}, 159 (1999). 

\bibitem{EOS:00}
%
J. B. Elliott et al., (EOS Collaboration), Phys. Rev. C {\bf 62}, 064603 (2000). 

\bibitem{VanHove}
L. Van Hove, Physica {\bf 15}, 951 (1949).

\bibitem{VanHove2}
%
L. Van Hove,   Physica {\bf 16},   137 (1950).

\bibitem{MyEOS:93}
%
M. I. Gorenstein, D. H. Rischke, H. Stocker, W. Greiner and 
K. A. Bugaev, 
J. Phys. G {\bf 19}, L69  (1993).


\bibitem{Kfactor:1}
%
D. Vretenar, T. Niksic, and P. Ring, Phys. Rev. C {\bf 68}, 024310 (2003). 


\bibitem{Kfactor:2}
%
G. Colo and Nguyen Van Giai, Nucl. Phys. A {\bf 731}, 15 (2004). 


\bibitem{Kfactor:3}
%
V. B. Soubbotin, V. I. Tselyaev and  X. Vinas, 
Phys. Rev. C {\bf  69}, 064312 (2004). 

\bibitem{Khan:2009}
%
E. Khan, Phys. Rev. C {\bf 80}, 011307(R) (2009). 


\bibitem{Stanley:71}
%
see, for instance, 
H. E. Stanley, {\it Introduction to phase transitions and critical phenomena} 
(Clarendon Press, Oxford, 1971).



\bibitem{Fisher:67}
M. E. Fisher, Physics {\bf 3},  255 (1967).

\bibitem{QGBSTM1}
%
K. A. Bugaev,
Phys. Rev. {\bf C 76},    014903 (2007); and
Phys. Atom. Nucl. {\bf 71},  1615 (2008) .

\bibitem{QGBSTM2}
%
K. A. Bugaev, V. K. Petrov and G. M. Zinovjev,
Phys.  Part. Nucl. Lett. {\bf 9}, 238  (2012);
arXiv:0904.4420  [hep-ph] (2009).

\bibitem{FWM:08}
%
K. A. Bugaev, V. K. Petrov and G. M. Zinovjev, Europhys. Lett.
{\bf 85},  22002 (2009); 
Phys. Rev.  {\bf C  79},  054913    (2009).

\bibitem{Aleksei:11}
%
A. I. Ivanytskyi,
Nucl. Phys. A {\bf 880}, 12 (2012). 

\bibitem{String:10}
%
K. A. Bugaev  and G. M. Zinovjev,
Nucl. Phys. A {\bf 848},   443 (2010).


\bibitem{String:11}
%
K. A. Bugaev, A. I. Ivanitskii, E. G. Nikonov, V. K. Petrov,  A. S. Sorin and G. M. Zinovjev,
 Phys. Atom. Nucl. {\bf 75},   707 (2012).


\bibitem{KABScaling:06}
%
J. B. Elliott, K. A. Bugaev, L. G. Moretto   and L.  Phair,
arXiv:0608022 [nucl-ex].

\bibitem{KABJGross:09}
%
J. Gross, 
J. Chem. Phys.   {\bf 131}, 204705  (2009).

\bibitem{SMM:12}
%
N. Buyukcizmeci et al.,  arXiv:1211.5990v2  [nucl-th]. 

\bibitem{Wilk:2008} 
  G.~Wilk and Z.~Wlodarczyk,
  Eur.\ Phys.\ J.\ A {\bf 40}, 299 (2009).

\bibitem{Frankland:2004}
  J.~D.~Frankland {\it et al.}  [INDRA and ALADIN Collaborations],
  Phys.\ Rev.\ C {\bf 71},   034607 (2005).
\end{thebibliography}
\end{document}